\ifx \kindle \undefined
\documentclass[]{article}
\usepackage[left=0.75in,top=1.125in,bottom=1.125in,right=0.75in,nohead]{geometry}
\else
\documentclass[fontsize=8pt]{scrartcl}
\usepackage[paperwidth=9cm, paperheight=12cm, textheight=4cm, top=1cm, left=1cm, right=1cm, bottom=1cm]{geometry}
\fi

\author{Stuart B. Heinrich \\ sbheinri@ncsu.edu }
\title{Physical Relativism as an Interpretation of Existence}

\usepackage[numbers,sort]{natbib}
\usepackage{xcolor,graphicx}
\usepackage[linkbordercolor=white,bookmarks=true]{hyperref} 
\usepackage{amssymb}
\usepackage[cmex10]{amsmath}
\usepackage{amsthm} 

\usepackage{verbatim}
\usepackage{url}
\usepackage{cancel}
\usepackage{color}
\usepackage{rotating}
\usepackage{multicol}
\usepackage{boxedminipage}
\usepackage{placeins} 
\usepackage{booktabs} 
\usepackage{times} 
\usepackage{dialogue}
\usepackage{lips} 
\usepackage{amsthm}

\newtheorem{theorem}{Theorem}

\newcommand{\eqnref}[1]{ (\ref{#1})}

\newcommand{\secref}[1]{Section \ref{#1}}
\newcommand{\thmref}[1]{Theorem \ref{#1}}

\newcommand{\argmax}{\operatornamewithlimits{argmax}}

\begin{document}
\maketitle

\begin{abstract}
Despite the success of modern physics in formulating mathematical theories that can predict the outcome of quantum-scale experiments, the physical interpretations of these theories remain controversial.  In this manuscript, we propose a new interpretation of existence that we call physical relativism.  Under physical relativism, the difference between mathematical existence and physical existence is clarified, and Wheeler's `it from bit' viewpoint can be objectively evaluated.  In addition, physical relativism provides a simple answer to the question of why the universe exists at all, and permits us to derive the maximally biophilic principle, a generalization of the anthropic principle that ascribes high prior likelihood to the observation of a universe with simple physical laws supporting the overall concepts of time, space and the emergent evolution of life.
\end{abstract}

\ifx \kindle \undefined
\begin{multicols}{2}
\else
\thispagestyle{empty}
\pagestyle{empty}
\fi

\section{Introduction}

Over the course of modern history, we have seen advances in biology, chemistry, physics and cosmology that have painted an ever-clearer picture of the physical processes explaining \emph{how} we came to exist in this universe.  However, despite all these advances, it seems we have not made any actual progress towards answering (with any certainty) the fundamental questions of `\emph{why}?' and `\emph{what is existence?}'

Ever since the probing of quantum-scale physics, scientific interpretations of reality have become divided over these fundamental questions, with interpretations ranging from the classical stochastic and observer-centric Copenhagen interpretation, to the deterministic and observer-free view promoted by de Broglie and Bohm, now known as Bohmian mechanics \citep{Pladevall12}, to the stochastic and observer-free theory of Ghirardi-Rimini-Weber \citep{Ghirardi86}, to various Many Worlds interpretations \citep{Vaidman02online}, such as Ultimate Ensemble theory \citep{Tegmark08} and even to Wheeler's solipsist `it from bit' viewpoint \citep{Wheeler90}.

In other words, scientific opinion has apparently become a matter of philosophy.  Therefore, we begin by introducing some of the basic philosophical perspectives.  In 510 BCE, Parmenides reasoned that \emph{ex nihilo nihil fit}, or ``nothing comes from nothing,'' meaning that the universe in the now implies an eternal universe without any specific moment of creation.  This viewpoint was shared by later Greek philosophers such as Aristotle and Plato, but does not really answer the question.  In 1697, \citet{LEIBNIZ1697} asked for ``a full reason why there should be any world rather than none.''  He claimed \citep{LEIBNIZ1714} that ``nothing takes place without sufficient reason,'' now known as the Principle of Sufficient Reason (PSR), and generalized the earlier question by asking, ``why is there something, rather than nothing?'', now known as the Primordial Existence Question (PEQ).

This fundamental question, further reviewed in \citet[p.296-301]{Edwards67} and \citet{Lutkehaus99}, has been echoed by many modern philosophers such as Richard Swinburne, who said, ``It remains to me, as to so many who have thought about the matter, a source of extreme puzzlement that there should exist anything at all'' \citep[p.283]{Swinburne91}, and Derek Parfit, who asked, ``Why is there a Universe at all? It might have been true that nothing ever existed; no living beings, no stars, no atoms, not even space or time. When we think about this possibility, it can seem astonishing that anything exists'' \citep[p.24]{Parfit98b}.

Most physicists and cosmologists are equally perplexed.  Richard Dawkins has called it a ``searching question that rightly calls for an explanatory answer'' \citep[p.155]{Dawkins06}, and Sam Harris says that ``any intellectually honest person will admit that he does not know why the universe exists.  Scientists, of course, readily admit their ignorance on this point'' \citep[p.74]{Harris06}.

With that said, modern inflationary cosmology does offer some powerful insights into these questions.  A generic property of inflation is that the universe began from a small quantum fluctuation \citep{Stenger90,Hartle83,Hawking98,Guth07,Guth97book,Pasachoff03} \citep[p.129]{Hawking88}\citep[p.131]{Hawking12}.  According to \citet{Vilenkin83}, ``A small amount of energy was contained in that [initial] curvature, somewhat like the energy stored in a strung bow. This ostensible violation of energy conservation is allowed by the Heisenberg uncertainty principle for sufficiently small time intervals. The bubble then inflated exponentially and the universe grew by many orders of magnitude in a tiny fraction of a second.''

According to Stephen Hawking, ``When one combines the theory of general relativity with quantum theory, the question of what happened before the beginning of the universe is rendered meaningless'' \citep[p.135]{Hawking12}, because, ``when we add the effects of quantum theory to the theory of relativity, in extreme cases warpage can occur to such an extent that time behaves like another dimension of space.  In the early universe--when the universe was small enough to be governed by both general relativity and quantum theory--there were effectively four dimensions of space and none of time'' \citep[p.134]{Hawking12}.

The notion that our timelike dimension emerged out of a spatial dimension is certainly controversial, and it is even more controversial when Hawking argues that, ``The realization that time behaves like space presents a new alternative.  It not only removes the age-old objection to the universe having a beginning, but also means that the beginning of the universe was governed by the laws of science and doesn't need to be set in motion by some God'' \citep[p.135]{Hawking12}, adding, ``Because there is a law like gravity [and quantum physics], the universe can and will create itself from nothing.  Spontaneous creation is the reason there is something rather than nothing, why the universe exists, why we exist'' \citep[p.180]{Hawking12}.  In other words, Hawking believes that Leibniz's question has been answered.

The flaw with this logic is that even if the mathematics of spontaneous creation are correct, they are based on the axioms of general relativity and quantum physics, which are not ``nothing.''  Thus, it is not the \emph{ex nihilo} creation of something from nothing, but rather, the derivation of a set of statements describing the universe in the now from a set of axioms.  We should not be surprised to learn that a set of statements can be derived from a set of axioms, because for \emph{any} set of non-contradictory statements, one could find an axiomatic system that derives all those statements simply by taking the statements themselves as axioms.  If Hawking is correct that the universe can be derived from the axioms of M-theory, then this is clearly a much reduced set of axioms, but he has done nothing to answer the question of \emph{why} those axioms are true, nor has he shown that they are the most fundamental possible set of axioms.  Thus, it leaves Leibniz' question completely untouched.

Most physicists do recognize this issue.  Brian Greene specifically pointed out that modern inflationary cosmology cannot resolve Leibniz's question \citep[p.310]{Greene04}, adding, ``If logic alone somehow required the universe to exist and be governed by a unique set of laws with unique ingredients, then perhaps we'd have a convincing story.  But to date, that's nothing but a pipe dream'' \citep[p.310]{Greene04}.

A theory that very nearly meets Greene's goal was proposed by \citet{Tegmark98}, known as the Mathematical Universe Hypothesis (MUH) or Ultimate Ensemble theory \citep{Tegmark03,Tegmark08}.  As formulated by Tegmark, the MUH rests on the sole postulate that ``all structures that exist mathematically also exist physically.''  The MUH is attractive because it permits a broader application of anthropic reasoning, but does not fully resolve Leibniz's question because one can still ask why this postulate is true.  It also does not clearly define the difference between mathematical and physical existence.

In this paper, we present logical arguments (\secref{sec_logic}) in support of a new interpretation of existence that we call \emph{physical relativism}.  We begin with a bottom-up logical argument (\secref{sec_axiomatization}), and then show how physical relativism answers Leibniz's question and resolves the paradox of infinite regress (\secref{sec_anthropic}), and finally show that physical relativism permits us to generalize the anthropic principle and predict, with high likelihood, many previously unexplained aspects about the observed laws of physics.  

Next, we refute some common criticisms of physical relativism (\secref{sec_refutations}); in particular, we show that physical relativism is compatible with G\"{o}del's theorems (\secref{sec_incomplete}) and observations of quantum randomness (\secref{sec_quantum_random}).  Finally, we discuss philosophical interpretations (\secref{sec_interp}) and give closing remarks (\secref{sec_conclusion}).

\section{Logical Arguments} \label{sec_logic}

The first argument (\secref{sec_axiomatization}) shows that if the universe has finite information content, then self-awareness can exist in abstract axiomatic systems, which implies physical relativism.  The second argument (\secref{sec_anthropic}) shows that anthropic reasoning is capable of resolving the paradox of infinite regress and simultaneously answering Leibniz's question, but \emph{only} under the premise of physical relativism.  In the third argument (\secref{sec_mbp}), we show that under physical relativism, the anthropic principle can be generalized to obtain the Maximally Biophilic Principle (MBP), and the MBP explains some of the most basic properties of our universe such as the concepts of causality, and approximate locality, with high likelihood.

\subsection{The Axiomatization Argument} \label{sec_axiomatization}

Any finite system can be formalized into an axiomatic system, for example by using one axiom to assert the truth of each independent piece of information.  Thus, assuming that our reality has \emph{finite information} content (\secref{sec_infinite}), there must be an axiomatic system that is isomorphic to our reality, where every true or false statement about reality can be proved as a theorem from the axioms of that system, and conversely any theorem of that system corresponds to a truth about reality.

By the principle of explosion, the presence of a single contradiction in an axiomatic system permits any statement to be proven true no matter how nonsensical \citep[p.18]{Franzen05}.  Some statements about reality are false, so it must also be possible to prove these corresponding statements formulated in the language of the axiomatic system are false, and thus we may conclude that the axiomatic system of our reality is \emph{consistent}.  Note that this conclusion is perfectly compatible with G\"{o}del's theorems (\secref{sec_incomplete}).

Self-aware life forms exist in our reality.  Thus, by the isomorphic property, it must be possible to derive self-awareness as a theorem from the (finite and consistent) axiomatic system of our reality.  

For any theorem of an axiomatic system, there must be other axiomatic systems that can also derive that theorem.  For example, a new axiomatic system can be found by the simple inclusion of a new axiom that does not contradict any existing axioms, or by modifying or subtracting an axiom that was not directly used in the derivation of the theorem.  Indeed, there are an infinite number of ways to modify an axiomatic system while keeping any particular theorem intact.  Thus, there must be an infinite number of different axiomatic systems that also derive the experience of self-awareness.

Still, we take for granted that our self-awareness and perception of a consistent reality is some kind of evidence that the axiomatic system isomorphic to our reality is somehow objectively unique in the sense that it is the \emph{only} axiomatic system that is manifested by a `real physical universe' or `reality.'  However, as we will now show, this assumption leads to contradiction:

\begin{proof}
\begin{quote} Let $\theta$ and $\phi$ be two different consistent axiomatic systems.  Assume that $\theta$ is objectively special in that it is manifested by a `real physical universe,' whereas $\phi$ is a purely theoretical system that has no physical manifestation.
\end{quote}

If it is objectively true that $\theta$ is manifested but $\phi$ is not, then this statement must be objectively derivable.  Because no axiomatic system can derive its own consistency, this derivation must be made from an external axiomatic system $\xi$.  If $\xi$ is a non-empty set, then it is not truly objective, but if $\xi$ is the empty set, then nothing can be derived from it.  Thus, by contradiction, our assumption that there is just one axiomatic system that is objectively special must be false.
\end{proof}

\begin{theorem} \label{proof_no_objective}
Either all consistent axiomatic systems correspond to `real physical universes,' or none of them do.
\end{theorem}

The fact that self-aware observers in our universe perceive a reality implies, by \thmref{proof_no_objective}, that either all axiomatic systems correspond to `real physical universes,' or that our own universe is not a `real physical universe.'  In case of the latter, physical manifestation must not be necessary for internally defined self-aware observers to perceive a reality.

Either way, the difference is purely semantical, because we are left with the conclusion that \emph{any consistent axiomatic system that derives self-aware observers is perceived as a reality by those self-aware observers}.  This is what we mean by \emph{physical relativism}.

\subsection{The Anthropic Argument} \label{sec_anthropic}

Habitability of a planet depends on a confluence of factors ranging from parent star class \citep{Kasting93,Kasting97} and stellar variation \citep{Lammer07}, to planet mass \citep{Raymond07}, composition, orbit distance \citep{Huggett95}, stability \citep{Lasker93}, early geochemistry conditions \citep{Parkinson08} and many other factors \citep{Irwin01}.  If all of these properties were chosen at random, without any overall guiding influence or purpose, then the probability of achieving conditions amenable to life must be exceedingly small.  Moreover, even on a theoretically habitable world, when we consider the probability of random chemistry interactions giving rise to self-replication and the actual evolution of life, our assessment of the overall probability of any random planet harboring life becomes even more diminutive.

Without knowing the precise details of all the chemical interactions that took place in order to give rise to life on Earth, it is difficult to make accurate predictions as to how small this probability actually is.  However, despite recent observations of potentially habitable exoplanets \citep{Borucki11} and better models of early biochemistry indicating that life might not be quite as rare as originally believed \citep[p.47]{Kauffman95}, it is only by taking into account our cosmological observations of billions upon billions of other star systems that we can explain the presence of life as something to be truly expected.

By reasoning of the anthropic principle that ``conditions observed must allow the observer to exist,'' it is truly only necessary for at least one of the practically infinite number of planets in the universe to contain life in order for us to resolve the mystery of why, when we look around, we should observe a planet with all the right conditions for life \citep{Penrose91}.

However, the mystery is still not fully solved, because it merely illustrates the remarkable tuning of the underlying laws of physics that permit a universe with the capacity for life.  From the molecular properties of water \citep{Henderson1913} to the precise balance between the strength of fundamental forces \citep{Dicke61}, to the number of dimensions and the precise values of all the fundamental constants, all of which exist in a perfect balance.

As stated by Paul Davies, ``There is now broad agreement among physicists and cosmologists that the universe is in several respects ‘fine-tuned' for life \citep{Davies03}.''  According to Stephen Hawking, ``The laws of science, as we know them at present, contain many fundamental numbers, like the size of the electric charge of the electron and the ratio of the masses of the proton and the electron...and the remarkable fact is that the values of these numbers seem to have been very finely adjusted to make possible the development of life'' \citep[p.125]{Hawking88}.  For example, if the strength of the strong nuclear force were changed by a mere 2\%, the physics of stars would be drastically altered so much that all the universe's hydrogen would have been consumed during the first few minutes after the big bang \citep[p.70-71]{Davies93}.

Can we again invoke the anthropic principle in order to answer the question of why the fundamental constants are fine-tuned to support life?  This second application is known as the ``strong'' anthropic princple (SAP) \citep{Barrow88}.  However, the SAP requires some evidence that there are an extremely large number (perhaps infinite) of different universes having different parameters for the physical constants.

It is believed by some that the modern incarnation of superstring theory known as M-theory \citep{Duff96} satisfies this condition.  Under M-theory, there are 11 dimensions of spacetime, 7 of which have been curled up into some Calabi-Yau manifold \citep{Candelas85}.  There are at least $10^{500}$ different ways to curl up these dimensions \citep[p.118]{Hawking12}, and each way would yield a universe with different fundamental constants \citep[p.372]{Greene04}.  It has been suggested that perhaps these correspond to parallel universes within a multiverse \citep[p.93]{Kaku05}, or as ``parallel histories'' of this universe \citep[p.136]{Hawking12}.  Either way, the selective power of the SAP could then explain why we exist in a universe with a particular set of fundamental constants amenable to life.

However, this is not the end of the story -- even if all the configurations allowed by M-theory were manifested in some kind of multiverse, one could still just as easily ask why the basic axioms of M-theory had been miraculously selected in order to give rise to a multiverse containing a universe capable of supporting life.  As noted by Greene, ``Even if a cosmological theory were to make headway on this question, we could ask why that particular theory--its assumptions, ingredients, and equations--was relevant, thus merely pushing the question of origin one step further back'' \citep[p.310]{Greene04}.  This paradox of infinite regress has been pondered since antiquity \citep[p.38]{Franzen05}.

Physical relativism says that \emph{all} consistent axiomatic systems are on equal grounds, and hence the anthropic principle is empowered to select from the infinite set of all consistent axiomatic systems, without requiring some arbitrary set of axioms as a starting point.  Thus, physical relativism solves the paradox of infinite regress, and simultaneously answers Leibniz's question in the simplest possible way: objectively speaking, nothing actually exists.

\subsection{The Maximally Biophilic Principle} \label{sec_mbp}

It was noted by \citet{Davies03} that the anthropic principle ``fails to distinguish between minimally biophilic universes, in which life is permitted but only marginally possible, and optimally biophilic universes, in which life flourishes because biogenesis occurs frequently''.  We now show that, under the premise of physical relativism, the logic behind the anthropic principle can be generalized into a more powerful principle that does select ``optimally biophilic'' universes over ``minimally biophilic'' ones, and this yields explanations to some very fundamental questions about the universe.

We first demonstrate by example.  Suppose that there are just two axiomatic systems that define self-aware life, one of which exists in a minimally biophilic universe harboring just a single self-aware life form, and the other being an `optimally biophilic' universe harboring 1 million self-aware life forms.  For any given self-aware being, without any prior knowledge, the likelihood that this being is from the optimally biophilic universe is $1000000/1000001$, or $99.9999\%$.

This logic can be formalized to prove that, in general, the axiomatic system that describes an observer's reality with greatest likelihood is the axiomatic system that defines the most self-aware observers.  Let $\Phi$ be the infinite set of all consistent axiomatic systems, and $\mathbb{S}$ be the (presumably infinite) set of self-aware observers defined by all axiomatic systems, and $S(\theta)$ be the set of self-aware beings derived by an axiomatic system $\theta \in \Phi$:

\begin{align}
S(\theta) = \{ s | \theta \vdash s \land s \in \mathbb{S} \}
\end{align}

The likelihood of a model given an event (equal to the probability of an event given a model) is the ratio of the number of cases favorable to it, to the total number of cases possible.  Thus, given an observer $s \in S(\theta')$, and without any additional prior knowledge, the likelihood of an axiomatic system $\theta$ being the observer's axiomatic system is given by:

\begin{equation}
\mathcal{L}(\theta = \theta' | s \in S(\theta') ) = \frac{ \#( S(\theta) ) }{ \sum_{\phi \in \Phi} \#( S(\phi) )  } \label{eq_likelihood}
\end{equation}

The anthropic principle states that `conditions of the observer must allow the observer to exist,' which means that if an axiomatic system $\theta$ does not define any self-aware observers, then the likelihood of $\theta$ being the observer's axiomatic system is zero:

\begin{equation}
\forall \theta \: \#( S(\theta) ) = 0 \implies \mathcal{L}(\theta = \theta' | s \in S(\theta') ) = 0
\end{equation}

Thus, the anthropic principle is trivially implied by \eqnref{eq_likelihood}.  However, this is not the most powerful statement we can make.  By taking the logarithm of \eqnref{eq_likelihood}, we see that the maximum likelihood estimate of the observer's axiomatic system is simply the system that defines the largest number of self-aware observers:

\begin{align}
\hat{ \theta }_{ML} &= \argmax_\theta \mathcal{L}(\theta| x_{\theta,s} ) \\
 &= \argmax_\theta \log \mathcal{L}(\theta| x_{\theta,s} ) \\
 &= \argmax_\theta \left\{ \log \#( S(\theta)) - \log \sum_{\phi \in \Phi} \#( S(\phi) ) \right\}\\
&= \argmax_\theta \#( S(\theta))
\end{align}

In other words, even before making any observations, physical relativism predicts that an observer should expect based on logic alone that his universe is an optimally biophilic one.  We call this the maximally biophilic principle (MBP).

Because physical relativism tells us that it is possible to derive self-aware observers axiomatically, it is conceivable that there are some axiomatic systems that derive just a single observer fairly directly.  However, an axiomatic system that instead indirectly derives self-aware beings via emergent processes such as evolution will be capable of deriving infinitely more self-aware beings with fewer axioms, and would thus have much higher likelihood according to the MBP.

It is impossible to have emergent processes without at least an approximate notion of causality, because without causality there can be no change.  It is also impossible to have emergent processes without at least some approximate notion of locality, because without spatial relationships there could be no shape, form, structure or complexity in the universe.  Thus, the MBP also implies high likelihood for axiomatic systems with spacelike and timelike dimensions, so we should not be surprised to observe those in our physics.

Finally, we should not be surprised to find that, at the smallest quantum scale, the universe is not \emph{perfectly} local or causal, because there is no difficulty in representing non-localities or temporal dependencies in axiomatic systems, and the MBP can only select for local and causal properties insofar as they permit the macroscopic capacity for emergent processes.  Moreover, if quantum phenomena are involved with the physics of consciousness as suggested by some recent research \citep{Jibu95,Penrose91,Vitiello01,Ritz04,Conte07,Conte09}, then this type of apparently non-deterministic and non-local behavior might be somehow required for the derivation of self-aware observers.

\section{Refutation of Common Objections} \label{sec_refutations}

A number of overall objections to the MUH have already been summarized and refuted by \citet{Tegmark08}.  Because the MUH is implied by physical relativism, many of refutations hold for physical relativism as well.  This section will mostly focus on refuting particular objections to the logical arguments presented in \secref{sec_logic}; most notably, Tegmark's own misgivings about G\"{o}del's theorems (\secref{sec_incomplete}).

\subsection{Incompatible with G\"{o}del's theorems?} \label{sec_incomplete}

Formally, an axiomatic system is called \emph{consistent} if it cannot prove any statement along with its negation (a contradiction), and \emph{complete} if every sentence that can be expressed in the language can be either proved or disproved.  G\"{o}del's first theorem shows that any axiomatic system containing a modicum of arithmetic power is incomplete, and his second theorem shows that any axiomatic system containing a modicum of arithmetic power cannot prove its own consistency \citep{GodelWorks86}.

These theorems have been the subject of many confusions and misunderstandings, as summarized in \citet{Franzen05}.  There is a commonly expressed fear that G\"{o}del's first theorem implies there will always be some truths about reality that cannot be proven, making it impossible to formulate a theory of physics that fully describes all aspects of reality.  For example, in his 2003 lecture at the Cambridge-MIT Institute (CMI), Stephen Hawking said, ``According to the positivist philosophy of science, a physical theory is a mathematical model.  So if there are mathematical results that cannot be proved, there are physical problems that cannot be predicted,'' adding, ``...some people will be very disappointed if there is not an ultimate theory, that can be formulated as a finite number of principles. I used to belong to that camp, but I have changed my mind'' \citep{Hawking03lecture}.

This sentiment was echoed by Freeman Dyson, who said, ``His theorem implies that pure mathematics is inexhaustible. No matter how many problems we solve, there will always be other problems that cannot be solved within the existing rules. Now I claim that because of G\"{o}del's theorem, physics is inexhaustible too'' \citep[p.225]{Freeman06}, and also Mark Alford, who said ``The methods allowed by formalists cannot prove all the theorems in a sufficiently powerful system [because of G\"{o}del's theorem].  The idea that math is `out there' is incompatible with the idea that it consists of formal systems'' \citep{Hut06}.

However, the first incompleteness theorem does \emph{not} imply that there will always be some truths about reality that cannot be proven \citep[p.24]{Franzen05}.  As pointed out in a response by Solomon Freeman and later conceded by Dyson, ``The basic equations of physics, whatever they may be, cannot indeed decide every arithmetical statement, but whether or not they are a complete description of the physical world, and what completeness might mean in such a case, is not something that the incompleteness theorem tells us anything about''\citep[p.88]{Franzen05}.  

The fundamental confusion arises from the false assumption that for every sentence that can be formulated in the language of a system, there must be some observation that an internal observer could make where the observed outcome is related to the decidability of the sentence.  In fact, it is fairly trivial to prove that this assumption is false (\thmref{proof_godel1}), which simply means that any indecidable statements that may exist simply have no relevance to the internally perceived reality.

\begin{proof}
\begin{quote} Assume there is an axiomatic system $\theta$ that defines the reality of a self-aware observer who has constructed an experiment with a single binary outcome that depends on the decidability of some indecidable statement $s$ that is written in the language of $\theta$.
\end{quote}

By definition, the statement $s$ is an indecidable statement of $\theta$, so the observer would not be able to derive the observed outcome of the experiment from $\theta$.  Thus, because $\theta$ fails to explain all aspects of the observers reality, contradiction is reached.
\end{proof}

\begin{theorem} \label{proof_godel1}
If a self-aware observer is defined by an incomplete axiomatic system, then it must be impossible for the observer to construct any experiment having an outcome that depends upon the decidability of an indecidable statement expressible in the language of that system.
\end{theorem}

\citet[p.21]{Tegmark08} has also expressed doubts with regards to G\"{o}del’s second theorem, lamenting that, ``Our standard model of physics includes everyday mathematical structures such as the integers (defined by the Peano axioms) and real numbers. Yet G\"{o}del’s second incompleteness theorem implies that we can never be 100\% sure that this everyday mathematics is consistent: it leaves open the possibility that a finite length proof exists within number theory itself demonstrating that 0 = 1.  Using this result, every other well-defined statement in the formal system could in turn be proven to be true and mathematics as we know it would collapse like a house of cards''.  In order to escape this issue, \citet{Tegmark08} proposed the more restricted Compute Universe Hypothesis (CUH) as an alternative to the MUH, which only includes axiomatic systems that are simple enough to escape these G\"{o}del-inspired worries.

However, Tegmark's fears in regard to the second theorem are also unfounded.  As explained by \citet[p.101]{Franzen05}, ``The second incompleteness theorem is a theorem about \emph{formal provability}, showing that...a \emph{consistent} theory T cannot postulate \emph{its own consistency}, although the consistency of T can be postulated in another consistent theory.''  In other words, an internal observer cannot prove the consistency of an axiomatic system that hypothesized to describe his reality by using the axioms of that system.

The fact that we cannot prove our theories are formally consistent, or prove that they are fully descriptive of the unobserved aspects of reality, does not preclude the existence of a consistent and fully descriptive axiomatic system that describes reality.  Indeed, this result is nearly identical to the way in which the Halting problem \citep[p.173]{Sipser06}, which shows us that one cannot write a finite length proof that any computer program will halt, does not preclude the existence of an arbitrarily long computer program that \emph{does} halt.

\subsection{Infinite Information?} \label{sec_infinite}

One potential argument against the axiomatization argument (\secref{sec_axiomatization}) is that the universe has an infinite information content, thereby preventing it from being represented by an axiomatic system.

Most cosmologists believe that there is a finite amount of energy in the observable universe, with recent analysis of 7-year data from the Wilkinson Microwave Anisotropy Probe (WMAP) estimating that of this finite amount, $72.8\% \pm -1.6\%$ is in the form of Dark Energy, $22.7\% \pm 1.4\%$ is in the form of Dark Matter, and $4.56\% \pm 0.16\%$ is in the form of regular baryonic matter \citep{Weiland11}.


However, this is just the observable universe, and we may still wonder if the universe has infinite extent.  According to the Friedmann-Lema\^{i}tre-Robertson-Walker (FLRW) model (or Standard model of cosmology \citep{Bergstrom06}), there are three possible overall `shapes' of the universe described by the curvature, $\Omega_k$, which can be deduced based on the density of matter.  

If $\Omega_{k} = 0$ exactly then the universe is flat and infinite, if $\Omega_k > 0$ then the universe is spherical and finite (and the curvature also tells us the size \citep{Milnor82}), and if $\Omega_k < 0$ then it is hyperbolic and infinite.  So far, the data has shown that the curvature is very \emph{close} to 0 \citep{Komatsu11}, but this is expected, and is insufficient to determine the sign \citep{Lesgourgues08}.  Indeed, if the magnitude of the true curvature is less than $10^{-4}$, then it might never be possible to determine by any future experiment \citep{Vardanyan09}.

Regardless, it is widely believed that the total positive energy of matter is exactly canceled out by the negative energy of gravity, thereby allowing the entire universe to be created out of the small amount of uncertainty in the vacuum energy of free space \citep[p.129]{Hawking88} \citep{Hartle83,Hawking98,Guth07,Guth97book,Pasachoff03} \citep[p.180]{Hawking12}.  If this is true, then the amount of positive energy is necessarily finite.

In addition, the Bekenstein bound \citep{Bekenstein08}, which can be derived from consistency between the laws of thermodynamics and general relativity \citep{Bekenstein81,Jacobson95,Bousso99,Bousso99b, Bekenstein00, Bousso02,Bousso03, Bekenstein05,Bekenstein08}, tells us that there is a finite information content in any finite region of space containing finite energy.    Thus, even if the universe did have infinite spatial extent and infinite energy (contrary to modern inflationary cosmology), there would still be a finite information content to any particular region of the universe (such as a galaxy).  This means that at the very minimum, an arbitrarily large finite region can be represented by an axiomatic system, and assuming that self-aware life can evolve in a finite region of space, then the logic of \secref{sec_axiomatization} would still hold in that region.

\subsection{Incompatible with Quantum Randomness?} \label{sec_quantum_random}

\citet[p.10]{Tegmark08} has claimed that the MUH is incompatible with true quantum randomness because it is impossible to generate a sequence of true random numbers using only axiomatic relationships.  While it is true that random numbers cannot be generated algorithmically, this does not preclude the existence of an axiomatic system that defines behavior which appears perfectly random based on the limited observations of an internal observer.

As a concrete example of this, consider the following set of axioms, which describe the position of a particle having position $X$ parameterized by integer-valued time $t$:

\begin{align*}
|| X(t) - X(t+1) || &= 1 \\
X(0) &= 1 \\
X(1) &= 2 \\
X(2) &= 3 \\
X(3) &= 2 \\
 &\vdots
\end{align*}

Suppose that this axiomatic system also defines an observer who, at time $t=2$, attempts to formulate a law describing the position of this particle as a function of time based on his observations for time $t \leq 2$.

It is clearly impossible for the observer to predict with certainty that $X(3) = 2$.  However, he might come up with the following theory:

\begin{quote}
 ``If a particle is observed at $X(t)$, then $X(t+1)$ will be uniformly randomly chosen from the set $\{ X(t-1), X(t+1) \}$.''
\end{quote}

It is evident that one could construct axioms such that this theory is validated to be an accurate description of the particle behavior, and perhaps the best possible description of particle behavior, over some arbitrary number of observations, despite that the axiomatic system itself has no notion of randomness.  In this case, the probability represents the observer's fundamental uncertainty in being able to predict certain axioms of the system.  Thus, the observed randomness in quantum physics is not incompatible with the notion that our reality is described by an axiomatic system.

\section{Interpretations} \label{sec_interp}

According to physical relativism, the distinction between physical existence and mathematical existence -- or equivalently, the difference between reality and the abstract -- is purely a point of perspective.  The difference is not that physical existence is a manifestation from potential truth to objective truth, but rather, that physical existence is the set of things that exist in the same mathematical universe as the observer.

A common philosophical question is whether or not our universe exists in an objective sense without the presence of self-aware observers, but this type of question neglects to recognize that the observer is an inextricable part of the universe.  The fact that our self-aware thoughts are capable of controlling our physical bodies is already proof that our thoughts are a part of the physics of our universe.  Because physical relativism holds that all aspects of reality (including self-aware thoughts) are defined by the same axiomatic system, there is no limit to the way these axioms may be intertwined and interdependent.  

We should, therefore, not be surprised but rather \emph{encouraged} to find a dependence on observers in quantum physics, because no theory of physics that fails to describe our thoughts and self-awareness would be complete, and it shows our physics is on the right track towards unification with a theory that describes self-aware thought.

To ask whether or not the universe exists in an objective sense without observers is thus akin to asking whether or not the universe would exist without gravity or electromagnetism.  We cannot simply remove fundamental axioms from the system and then say, `yes, it still exists.'  However, if we say `no, the universe does not exist without observers,' that should not be viewed as a trivialization of reality in the solipsist sense, either.  Rather, the people who ask this question are inadvertently trivializing the degree to which observers play in reality.

\citet{Wheeler90} believed that the physical world was a figment of the imagination, and that everything physical derives its existence from the observations made by observers, saying, ``... every it--every particle, every field of force, even the spacetime continuum itself--derives its function, its meaning, its very existence entirely--even if in some contexts indirectly--from the apparatus-elicited answers to yes-or-no questions, binary choices, bits.''

This is not the conclusion of physical relativism, because physical relativism asserts that there is no form of existence more objective than mathematical existence, and mathematical existence is not dependent on the presence of observers, nor is it created within the imagination of an observer.  Nonetheless, we may agree with Wheeler in the sense that the physical world derives its \emph{meaning and significance} from the observations we make.  We might even, perhaps, agree that \emph{physical existence} is derived from our observations, but only in the sense that we defined the term physical existence relative to the observer.

With regard to Tegmark's Mathematical Universe Hypothesis (MUH) that ``all structures that exist mathematically also exist physically'', the implication is that from an objective standpoint, mathematical existence is equivalent to physical existence.  Thus, according to Tegmark, different contradictory physical universes may exist, and hence physical existence is relative.  In other words, the MUH is objectively equivalent to physical relativism, although we have defined `physical existence' as a relative term, whereas Tegmark defined it absolutely. 

\section{Conclusion} \label{sec_conclusion}

Physical relativism was derived as a logical constraint on the philosophy of existence, based on the single well-supported assumption that the universe has finite information.

Physical relativism tells us that self-aware observers can exist in axiomatic systems without objective manifestation, and that the distinction between physical existence and mathematical existence is merely a point of perspective: physical existence is how self-aware observers describe things that exist in the same mathematical universe as themselves.

Physical relativism is not experimentally verifiable in the classical sense because it does not make any new measurable predictions.  Rather, it predicts that which we have already observed and struggled to explain: it predicts why the observed universe exists in a state that \emph{is} highly tuned to be flourishing with life, and answers the question of why a universe exists at all.

\bibliographystyle{plainnat}
\bibliography{reality}

\begin{thebibliography}{70}
\providecommand{\natexlab}[1]{#1}
\providecommand{\url}[1]{\texttt{#1}}
\expandafter\ifx\csname urlstyle\endcsname\relax
  \providecommand{\doi}[1]{doi: #1}\else
  \providecommand{\doi}{doi: \begingroup \urlstyle{rm}\Url}\fi

\bibitem[Barrow and Tipler(1988)]{Barrow88}
John~D. Barrow and Frank~J. Tipler.
\newblock \emph{The Anthropic Cosmological Principle}.
\newblock Oxford University Press, Oxford, 1988.

\bibitem[Bekenstein(1981)]{Bekenstein81}
Jacob Bekenstein.
\newblock Universal upper bound on the entropy-to-energy ratio for bounded
  systems.
\newblock \emph{Physical Review D}, 23\penalty0 (2):\penalty0 287--298, 1981.

\bibitem[Bekenstein(2005)]{Bekenstein05}
Jacob Bekenstein.
\newblock How does the entryop/information bound work?
\newblock \emph{Foundations of Physics}, 35\penalty0 (11):\penalty0 1805--1823,
  2005.

\bibitem[Bekenstein(2008)]{Bekenstein08}
Jacob Bekenstein.
\newblock Bekenstein bound.
\newblock \emph{Scholarpedia}, 3\penalty0 (10):\penalty0 7374, 2008.

\bibitem[Bekenstein(2000)]{Bekenstein00}
Jacob~D. Bekenstein.
\newblock Holographic bound from second law of thermodynamics.
\newblock \emph{Physics Letters B}, 481\penalty0 (2-4):\penalty0 339--345,
  2000.

\bibitem[Bergstr\"{o}m and Goobar(2006)]{Bergstrom06}
L.~Bergstr\"{o}m and A~Goobar.
\newblock \emph{Cosmology and Particle Astrophysics}.
\newblock Sprint, 2006.

\bibitem[Borucki et~al.(2011)Borucki, Koch, et~al.]{Borucki11}
William~J. Borucki, David~G Koch, et~al.
\newblock Characteristics of planetary candidates observed by kepler, ii:
  Analysis of the first four months of data.
\newblock 2011.

\bibitem[Bousso(1999{\natexlab{a}})]{Bousso99}
Raphael Bousso.
\newblock A covariant entropy conjecture.
\newblock \emph{Journal of High Energy Physics}, 17\penalty0 (4),
  1999{\natexlab{a}}.

\bibitem[Bousso(1999{\natexlab{b}})]{Bousso99b}
Raphael Bousso.
\newblock Holography in general space-times.
\newblock \emph{Journal of High Energy Physics}, 1999\penalty0 (6),
  1999{\natexlab{b}}.

\bibitem[Bousso(2002)]{Bousso02}
Raphael Bousso.
\newblock The holographic principle.
\newblock \emph{Reviews of Modern Physics}, 74\penalty0 (3):\penalty0 825--874,
  2002.

\bibitem[Bousso et~al.(2003)Bousso, Flanagan, and Marolf]{Bousso03}
Raphael Bousso, Eanna Flanagan, and Donald Marolf.
\newblock Simple sufficient conditions for the generalized covariant entropy
  bound.
\newblock \emph{Physical Review D}, 68\penalty0 (6), 2003.

\bibitem[Candelas et~al.(1985)Candelas, Horowitz, Strominger, and
  Witten]{Candelas85}
Philip Candelas, Gary Horowitz, Andrew Strominger, and Edward Witten.
\newblock Vacuum configurations for superstrings.
\newblock \emph{Nuclear Physics B}, 258:\penalty0 46--74, 1985.

\bibitem[Conte et~al.(2007)Conte, Todarello, Federici, Vitiello, Lopane,
  Khrennikov, and Zbilut]{Conte07}
Elio Conte, Orlando Todarello, Antonio Federici, Francesco Vitiello, Michele
  Lopane, Andrei Khrennikov, and Joseph~P. Zbilut.
\newblock Some remarks on an experiment suggesting quantum like behavior of
  cognitive entities and formulation of an abstract quantum mechanical
  formalism to describe cognitive entity and its dynamics.
\newblock \emph{Chaos, Solitons and Fractals}, 31:\penalty0 1076--1088, 2007.

\bibitem[Conte et~al.(2009)Conte, Khrennikov, Todarello, Federici,
  Mendolicchio, and Zbilut]{Conte09}
Elio Conte, Andrei~Yuri Khrennikov, Orlando Todarello, Antonio Federici,
  Leonardo Mendolicchio, and Joseph~P. Zbilut.
\newblock Mental states follow quantum mechanics during perception and
  cognition of ambiguous figures.
\newblock \emph{Open Systems and Information Dynamics}, 16\penalty0
  (1):\penalty0 1--17, 2009.

\bibitem[Davies(1993)]{Davies93}
Paul Davies.
\newblock \emph{The Accidental Universe}.
\newblock Cambridge University Press, 1993.

\bibitem[Davies(2003)]{Davies03}
Paul~C. Davies.
\newblock How bio-friendly is the universe?
\newblock \emph{International journal of astrobiology}, 2\penalty0
  (2):\penalty0 115--120, apr 2003.

\bibitem[Dawkins(2006)]{Dawkins06}
Robert Dawkins.
\newblock \emph{The God delusion}.
\newblock Houghton Mifflin, New York, 2006.

\bibitem[Dicke(1961)]{Dicke61}
R.~H. Dicke.
\newblock Dirac's cosmology and mach's principle.
\newblock \emph{Nature}, 192\penalty0 (4801):\penalty0 440--441, nov 1961.

\bibitem[Duff(1996)]{Duff96}
M.J. Duff.
\newblock M-theory (the theory formerly known as strings).
\newblock \emph{International Journal of Modern Physics A}, 11:\penalty0
  5623--5642, 1996.

\bibitem[Dyson(2006)]{Freeman06}
Freeman Dyson.
\newblock \emph{The Scientist as Rebel}.
\newblock New York Review Books, New York, 2006.

\bibitem[Edwards(1967)]{Edwards67}
P.~Edwards.
\newblock \emph{The encyclopedia of philosophy}.
\newblock Macmillan Reference, 1967.

\bibitem[Franz\'{e}n(2005)]{Franzen05}
Torkel Franz\'{e}n.
\newblock \emph{G\"{o}del's Theorem: An Incomplete Guide to Its Use and Abuse}.
\newblock A K Peters, Wellesley, Massachusetts, 2005.

\bibitem[Ghirardi et~al.(1986)Ghirardi, Rimini, and Weber]{Ghirardi86}
G.~C. Ghirardi, A.~Rimini, and T.~Weber.
\newblock Unified dynamics for microscopic and macroscopic systems.
\newblock \emph{Phys. Rev. D}, 34:\penalty0 470--491, Jul 1986.

\bibitem[G\"{o}del(1986)]{GodelWorks86}
Kurt G\"{o}del.
\newblock \emph{Kurt G\"{o}del Collected Works, Volume 1: Publications
  1929-1936}.
\newblock Oxford University Press, Oxford, 1986.

\bibitem[Greene(2004)]{Greene04}
Brian Greene.
\newblock \emph{The Fabric of the Cosmos: Space, Time, and the Texture of
  Reality}.
\newblock Knopf, New York, 2004.

\bibitem[Guth(1997)]{Guth97book}
Alan~H. Guth.
\newblock \emph{The Inflationary Universe: Quest for a New Theory of Cosmic
  Origins}.
\newblock Jonathan Cape, 1997.

\bibitem[Guth(2007)]{Guth07}
Alan~H Guth.
\newblock Eternal inflation and its implications.
\newblock \emph{arXiv}, 2007.

\bibitem[Harris(2006)]{Harris06}
Sam Harris.
\newblock \emph{Letter to a Christian nation}.
\newblock A.A.Knopf, New York, 2006.

\bibitem[Hartle and Hawking(1983)]{Hartle83}
J.~B. Hartle and S.~W. Hawking.
\newblock Wave function of the universe.
\newblock \emph{Phys. Rev. D}, 28:\penalty0 2960--2975, Dec 1983.
\newblock \doi{10.1103/PhysRevD.28.2960}.
\newblock URL \url{http://link.aps.org/doi/10.1103/PhysRevD.28.2960}.

\bibitem[Hawking and Turok(1998)]{Hawking98}
S.~W. Hawking and Neil Turok.
\newblock Open inflation without false vacua.
\newblock \emph{Physics Letters B}, 425\penalty0 (1-2):\penalty0 25--32, 1998.

\bibitem[Hawking(1988)]{Hawking88}
Stephen Hawking.
\newblock \emph{A Brief History of Time}.
\newblock Bantam Books, 1988.

\bibitem[Hawking(2003)]{Hawking03lecture}
Stephen Hawking.
\newblock G\"{o}del and the end of physics.
\newblock University Lecture, 2003.

\bibitem[Hawking and Mlodinow(2012)]{Hawking12}
Stephen Hawking and Leonard Mlodinow.
\newblock \emph{The Grand Design}.
\newblock Bantam Books, 2012.

\bibitem[Henderson(1913)]{Henderson1913}
Lawrence~J. Henderson.
\newblock The fitness of the environment, an inquiry into the biological
  significance of the properties of matter.
\newblock \emph{The American Naturalist}, 47\penalty0 (554):\penalty0 pp.
  105--115, 1913.
\newblock URL \url{http://www.jstor.org/stable/2455869}.

\bibitem[Huggett(1995)]{Huggett95}
Richard Huggett.
\newblock \emph{Geoecology: An Evolutionary Approach}.
\newblock Routledge, London, 1995.

\bibitem[Hut et~al.(2006)Hut, Alford, and Tegmark]{Hut06}
Piet Hut, Mark Alford, and Max Tegmark.
\newblock On math, matter and mind.
\newblock \emph{Foundations of Physics}, 36:\penalty0 765--794, 2006.

\bibitem[Irwin and Schulze-Makuch(2001)]{Irwin01}
Louis~Neal Irwin and Dirk Schulze-Makuch.
\newblock Assessing the plausibility of life on other worlds.
\newblock \emph{Astrobiology}, 1\penalty0 (2):\penalty0 143--160, 2001.

\bibitem[Jacobson(1995)]{Jacobson95}
Ted Jacobson.
\newblock Thermodynamics of spacetime: The einstein equation of state.
\newblock \emph{Physical Review Letters}, 75\penalty0 (7):\penalty0 1260--1263,
  1995.

\bibitem[Jibu and Tasue(1995)]{Jibu95}
M.~Jibu and K.~Tasue.
\newblock \emph{Quantum Brain Dynamics: An Introduction}.
\newblock John Benjamins, Amsterdam, 1995.

\bibitem[Kaku(2005)]{Kaku05}
Michio Kaku.
\newblock \emph{Parallel Worlds}.
\newblock Doubleday, New York, 2005.

\bibitem[Kasting et~al.(1993)Kasting, Whitmore, and Reynolds]{Kasting93}
James~F. Kasting, Daniel~P. Whitmore, and Ray~T. Reynolds.
\newblock Habitable zones around main sequence stars.
\newblock \emph{Icarus}, 101:\penalty0 108--128, 1993.

\bibitem[Kasting et~al.(1997)Kasting, Whittet, and Sheldon]{Kasting97}
James~F. Kasting, D.C. Whittet, and W.R. Sheldon.
\newblock Ultraviolet radiation from f and k stars and implications for
  planetary habitability.
\newblock \emph{Origins of Life and Evolution of Biospheres}, 27\penalty0
  (4):\penalty0 413--320, 1997.

\bibitem[Kauffman(1995)]{Kauffman95}
Stuart Kauffman.
\newblock \emph{At Home in the Universe}.
\newblock Oxford University Press, New York, 1995.

\bibitem[Komatsu et~al.(2011)Komatsu, Smith, Dunkley, Bennett, Gold, Hinshaw,
  Jarosik, Larson, Nolta, Page, Spergel, Halpern, Hill, Kogut, Limon, Meyer,
  Odegard, Tucker, Weiland, Wollack, and Wright]{Komatsu11}
E.~Komatsu, K.~M. Smith, J.~Dunkley, C.~L. Bennett, B.~Gold, G.~Hinshaw,
  N.~Jarosik, D.~Larson, M.~R. Nolta, L.~Page, D.~N. Spergel, M.~Halpern, R.~S.
  Hill, A.~Kogut, M.~Limon, S.~S. Meyer, N.~Odegard, G.~S. Tucker, J.~L.
  Weiland, E.~Wollack, and E.~L. Wright.
\newblock Seven-year wilkinson microwave anisotropy probe (wmap) observations:
  Cosmological interpretation.
\newblock \emph{The Astrophysical Journal Supplement Series}, 192\penalty0
  (2):\penalty0 18, 2011.

\bibitem[Lammer et~al.(2007)Lammer, Lichtenegger, Kulikov, et~al.]{Lammer07}
Helmut Lammer, Herbert Lichtenegger, Yuri~N Kulikov, et~al.
\newblock Coronal mass ejection (cme) activity of low mass m stars as an
  important factor for the habitability of terrestrial exoplanets. ii.
  cme-induced ion pick up of earth-like exoplanets in close-in habitable zones.
\newblock \emph{Astrobiology}, 7\penalty0 (1):\penalty0 185--207, 2007.

\bibitem[Lasker et~al.(1993)Lasker, Joutel, and Robutel]{Lasker93}
J.~Lasker, F.~Joutel, and P.~Robutel.
\newblock Stabilization of the earth's obliquity by the moon.
\newblock \emph{Nature}, 361\penalty0 (6413):\penalty0 615--617, 1993.

\bibitem[Leibniz(1697)]{LEIBNIZ1697}
G.W. Leibniz.
\newblock On the ultimate origination of things.
\newblock Technical report, Publishers name, 1697.
\newblock Reprinted in G.H.R. Parkinson \& M. Morris, 1973,
  Leibniz:philosophical writings (pp.136-144). London: J.M. Dent \& Sons.

\bibitem[Leibniz(1714)]{LEIBNIZ1714}
G.W. Leibniz.
\newblock Principles of nature and of grace founded on reason.
\newblock Technical report, 1714.
\newblock Reprinted in G.H.R. Parkinson \& M. Morris, 1973,
  Leibniz:philosophical writings (pp.195-204). London: J.M. Dent \& Sons.

\bibitem[Lesgourgues et~al.(2008)Lesgourgues, Starbobinsky, and
  Valkenburg]{Lesgourgues08}
Julien Lesgourgues, Alexei~A. Starbobinsky, and Wessel Valkenburg.
\newblock What do wmap and sdss really tell about inflation?
\newblock \emph{Journal of Cosmology and Astroparticle Physics}, 2008\penalty0
  (1), 2008.

\bibitem[L\"{u}tkehaus(1999)]{Lutkehaus99}
Ludger L\"{u}tkehaus.
\newblock \emph{Nichts: Abschied vom Sein, Edge der Angst}.
\newblock Haffmans bei Zweitausendeins, Z\"{u}rich, 1999.

\bibitem[Milnor(1982)]{Milnor82}
John~W. Milnor.
\newblock Hyperbolic geometry: the first 150 years.
\newblock \emph{Bulletin of the American Mathematical Society}, 6\penalty0
  (1):\penalty0 9--24, 1982.

\bibitem[Parfit(1998)]{Parfit98b}
D.~Parfit.
\newblock Why anything? why this?
\newblock \emph{London Review of Books}, 20\penalty0 (2):\penalty0 24--27,
  1998.

\bibitem[Parkinson et~al.(2008)Parkinson, Liang, Yung, and
  Kirschivnk]{Parkinson08}
C.D. Parkinson, M.C. Liang, Y.L. Yung, and J.L. Kirschivnk.
\newblock Habitability of enceladus: Planetary conditions for life.
\newblock \emph{Origins of Life Evolution Biospheres}, 38\penalty0
  (4):\penalty0 355--369, 2008.

\bibitem[Pasachoff and Filippenko(2003)]{Pasachoff03}
Jay~M. Pasachoff and Alex Filippenko.
\newblock \emph{The Cosmos: Astronomy in the New Millenium}.
\newblock Brooks Cole, 2003.

\bibitem[Penrose(1991)]{Penrose91}
Roger Penrose.
\newblock \emph{The Emperor's New Mind}.
\newblock Penguin, 1991.

\bibitem[Pladevall and Mompart(2012)]{Pladevall12}
Xavier~Oriols Pladevall and Jordi Mompart.
\newblock \emph{Applied Bohmian Mechanics: From Nanoscale Systems to
  Cosmology}.
\newblock Pan Stanford Publishing, 2012.

\bibitem[Raymond et~al.(2007)Raymond, Quinn, and Lunine]{Raymond07}
Sean~N. Raymond, Thomas Quinn, and Jonathan Lunine.
\newblock High-resolution simulations of the final assembly of earth-like
  planets 2: water delivery and planetary habitability.
\newblock \emph{Astrobiology}, 7\penalty0 (1):\penalty0 66--84, 2007.

\bibitem[Ritz et~al.(2004)Ritz, Thalau, Phillips, Wiltschko, and
  Wiltschko]{Ritz04}
Thorsten Ritz, Peter Thalau, John~B. Phillips, Roswitha Wiltschko, and Wolfgang
  Wiltschko.
\newblock Resonance effects indicate a radical-pair mechanism for avian
  magnetic compass.
\newblock \emph{Nature}, 429:\penalty0 177--180, 2004.

\bibitem[Sipser(2006)]{Sipser06}
Michael Sipser.
\newblock \emph{Introduction to the Theory of Computation}.
\newblock PWS Publishing, Boston, Massachusetts, 2006.

\bibitem[Stenger(1990)]{Stenger90}
V.~J. Stenger.
\newblock The universe: the ultimate free lunch.
\newblock \emph{European Journal of Physics}, 11\penalty0 (4):\penalty0 236,
  1990.

\bibitem[Swinburne(1991)]{Swinburne91}
R.~Swinburne.
\newblock \emph{The existence of God}.
\newblock Oxford University Press/Clarendon Press, Oxford/New York, 1991.

\bibitem[Tegmark(1998)]{Tegmark98}
Max Tegmark.
\newblock Is `the theory of everything'' merely the ultimate ensemble theory?
\newblock \emph{Annals of Physics}, 270\penalty0 (1):\penalty0 1--51, 1998.

\bibitem[Tegmark(2003)]{Tegmark03}
Max Tegmark.
\newblock Parallel universes.
\newblock \emph{arXiv}, 2003.

\bibitem[Tegmark(2008)]{Tegmark08}
Max Tegmark.
\newblock The mathematical universe.
\newblock \emph{Foundations of Physics}, 38\penalty0 (2):\penalty0 101--150,
  2008.

\bibitem[Vaidman(2002)]{Vaidman02online}
Lev Vaidman.
\newblock The many-worlds interpretation of quantum mechanics, Mar 2002.
\newblock URL \url{http://plato.stanford.edu/entries/qm-manyworlds/}.

\bibitem[Vardanyan et~al.(2009)Vardanyan, Trotta, and Silk]{Vardanyan09}
Mihran Vardanyan, Roberto Trotta, and Joseph Silk.
\newblock How flat can you get? a model comparison perspective on the curvature
  of the universe.
\newblock \emph{Monthly Notices of the Royal Astronomical Society},
  397\penalty0 (1):\penalty0 431--444, 2009.

\bibitem[Vilenkin(1983)]{Vilenkin83}
Alexander Vilenkin.
\newblock Birth of inflationary universes.
\newblock \emph{Phys. Rev. D}, 27:\penalty0 2848--2855, Jun 1983.
\newblock \doi{10.1103/PhysRevD.27.2848}.
\newblock URL \url{http://link.aps.org/doi/10.1103/PhysRevD.27.2848}.

\bibitem[Vitiello(2001)]{Vitiello01}
Giuseppe Vitiello.
\newblock \emph{My Double Unveiled}.
\newblock John Benjamins, 2001.

\bibitem[Weiland et~al.(2011)Weiland, Odegard, Hill, Wollack, Hinshaw, Greason,
  Jarosik, Page, Bennett, Dunkley, Gold, Halpern, Kogut, Komatsu, Larson,
  Limon, Meyer, Nolta, Smith, Spergel, Tucker, and Wright]{Weiland11}
J.~L. Weiland, N.~Odegard, R.~S. Hill, E.~Wollack, G.~Hinshaw, M.~R. Greason,
  N.~Jarosik, L.~Page, C.~L. Bennett, J.~Dunkley, B.~Gold, M.~Halpern,
  A.~Kogut, E.~Komatsu, D.~Larson, M.~Limon, S.~S. Meyer, M.~R. Nolta, K.~M.
  Smith, D.~N. Spergel, G.~S. Tucker, and E.~L. Wright.
\newblock Seven-year wilkinson microwave anisotropy probe (wmap) observations:
  Planets and celestial calibration sources.
\newblock \emph{The Astrophysical Journal Supplement Series}, 192\penalty0
  (2):\penalty0 19, 2011.

\bibitem[Wheeler(1990)]{Wheeler90}
John~A. Wheeler.
\newblock \emph{{Information, physics, quantum: The search for links}}.
\newblock Physics Dept., Princeton University, 1990.

\end{thebibliography}

\ifx \kindle \undefined
\end{multicols}
\fi

\end{document}